\pdfoutput=1
\documentclass[conference]{IEEEtran}
\IEEEoverridecommandlockouts

\usepackage{amsmath,amsfonts}
\usepackage{amssymb}
\usepackage{algorithmic}
\usepackage{graphicx}
\usepackage{subcaption}
\usepackage{textcomp}
\usepackage{xcolor}
\usepackage{placeins}
\usepackage{algorithm}

\usepackage{booktabs} 
\usepackage{siunitx} 

\newcommand{\Prx}{P_{\mathrm{rx}}}
\newcommand{\Ptx}{P_{\mathrm{tx}}}
\newcommand{\PL}{PL}
\newcommand{\Nf}{N_{f}}
\newcommand{\Nought}{N_{0}}
\newcommand{\BW}{BW}
\newcommand{\NF}{NF_{\mathrm{rx}}}
\newcommand{\SNRth}{SNR_{\mathrm{th}}}
\newcommand{\SIRth}{SIR_{\mathrm{th}}}
\newcommand{\Nest}{N_{\mathrm{est}}}
\newcommand{\CWmin}{CW_{\mathrm{min}}}
\newcommand{\dzero}{d_0}
\newcommand{\ple}{\mathrm{n}}

\def\BibTeX{{\rm B\kern-.05em{\sc i\kern-.025em b}\kern-.08em
    T\kern-.1667em\lower.7ex\hbox{E}\kern-.125emX}}

\begin{document}

\title{Enhanced V2X Communication Using Game-Theory Based Adaptive MAC Protocols \\

}

\author{\IEEEauthorblockN{Dhrumil Bhatt\textsuperscript{*}}
\IEEEauthorblockA{\textit{Department of Electrical and
} \\
\textit{Electronics Engineering}\\
\textit{Manipal Institute of Technology}\\
\textit{Manipal Academy of Higher Education}\\
Manipal, India \\
dhrumil.bhatt@gmail.com}
\and

\IEEEauthorblockN{Nirbhay Singhal\textsuperscript{*}}
\IEEEauthorblockA{\textit{Department of Electrical and} \\
\textit{Electronics Engineering}\\
\textit{Manipal Institute of Technology}\\
\textit{Manipal Academy of Higher Education}\\
Manipal, India \\
nirbhaysinghal09@gmail.com}

\thanks{\textsuperscript{*}These authors contributed equally to this work.}
}
\maketitle

\begin{abstract}
This paper presents an enhanced Vehicle-to-Everything (V2X) communication system featuring adaptive Medium Access Control (MAC) using game theory. Our approach integrates dynamic transmission power control, dynamic beacon rates, contention window adaptation, and implicit acknowledgment mechanisms within a Manhattan-like grid-based mobility scenario.  Simulations are conducted in a circular coverage area, incorporating refined signal propagation models and probabilistic vehicle mobility with boundary reflection. The results demonstrate effective beacon delivery with average delays under 0.35 s and packet loss rates less than 1\% in high-density conditions—specifically, with up to 80 vehicles operating within a 250 m radius. Key innovations include game theory-based environment-aware transmission parameter adaptation and a scalable design suited for interference-prone V2X deployments.
\end{abstract}

\section{Introduction}

The rapid evolution of Intelligent Transportation Systems (ITS) and autonomous driving technologies hinges on reliable, low-latency communication between vehicles and their environment. Vehicle-to-Everything (V2X) communication including Vehicle-to-Vehicle (V2V), Vehicle-to-Infrastructure (V2I), and Vehicle-to-Pedestrian (V2P) serves as the foundation of connected mobility \cite{ITS_Integration_Challenges, ITS_Cooperative_Intelligence}. By exchanging periodic beacons containing critical information such as vehicle position, speed, and trajectory, V2X facilitates cooperative perception, collision avoidance, and efficient traffic coordination \cite{ V2X_Enhancements_Coop_Driving}.

Despite its potential, the effectiveness of V2X communication beacon broadcasting is highly sensitive to wireless channel conditions, mobility patterns, and network load. In dense urban scenarios, the proliferation of simultaneous transmissions can lead to severe packet collisions and channel congestion, degrading delivery rates and increasing latency \cite{C-V2X_Multiple_Access_Congestion, SPS_Access_Collision_Analysis}. Additionally, simulation-based studies on C-V2X reveal that delivery reliability under Semi-Persistent Scheduling (SPS) is strongly influenced by factors like distance and the presence of hidden terminals \cite{SPS_Access_Collision_Analysis}. To analyze this behavior, analytical models have been developed to estimate access collision probabilities under 5G New Radio's SPS, demonstrating how MAC and PHY configurations impact performance.

Accurate signal-level simulation of the physical layer is computationally prohibitive for large-scale vehicular networks due to high mobility and density \cite{PHY_Abstraction_Methodology}. Consequently, many studies adopt simplified physical-layer abstractions by estimating SINR using node positions, accounting for path-loss, small-scale fading, and interference. While computationally efficient, such models risk overlooking key dynamics affecting real-world performance \cite{PHY_Abstraction_Methodology}.

Building on insights from these prior studies and acknowledging the associated challenges, we propose a V2X simulation framework that integrates:

\begin{itemize}
  \item Adaptive MAC strategies using dynamic contention window and transmit power control based on local vehicle density.
  \item An implicit acknowledgment mechanism to reduce redundant retransmissions.
  \item Game theory-based approach for variable beacon rates and transmission power. 
  
\end{itemize}

Although prior studies have extensively evaluated V2X MAC protocols using performance metrics such as collision probability and average delay \cite{C-V2X_vs_80211p_Markov}, average delay in cooperative MAC optimization \cite{Coop_MAC_RSU_Distributed}, and average throughput under semi-persistent scheduling schemes \cite{SPS_Throughput_5G_NR}. Our work places particular emphasis on beacon loss rate (BLR) and targeted delay measurements metrics, also considered in protocols like \cite{GNC-MAC} and \cite{NC-MAC}, as the system relies on Beacon Scheduling as opposed to traditional packet-based peer-to-peer networks.

In this context, \cite{NC-MAC} and \cite{GNC-MAC} are evaluated in terms of their BLR performance under comparable urban conditions. Additionally, \cite{Coop_MAC_RSU_Distributed} and \cite{efficient_coding} are analyzed concerning their overall MAC-layer strategies and adaptability to dense and dynamic environments. These protocols form a diverse and relevant baseline for assessing performance and the design alignment with our proposed approach.

\section{Related Work}

Improving beacon reliability in VANETs has driven research across MAC-layer strategies, including network coding, cooperative relaying, and adaptive access control.

\subsection{NC-MAC}

  Hamed Mosavat-Jahromi et al. propose a TDMA-based MAC protocol enhanced with network coding. Each vehicle transmits its own beacon or a coded linear combination of overheard messages during two allocated transmission opportunities. When a vehicle fails to receive a message, it broadcasts a NACK. Others that overhear both the original message and the NACK create coded packets, such as:

\begin{align}
C_1 &= \alpha_{11}m_1 + \alpha_{12}m_2 + \alpha_{13}m_3, \\
C_2 &= \alpha_{21}m_1 + \alpha_{22}m_2 + \alpha_{23}m_3,
\end{align}

allowing missed beacons to be recovered through matrix decoding. While efficient in highway scenarios, it assumes a 1D topology, relies on maintaining SINR-based neighbor lists, and requires multiple linearly independent packets to decode—limitations in mobile, broadcast-heavy 2D urban environments \cite{NC-MAC}.

\subsection{GNC-MAC}

Yu Li et al. expands on NC-MAC with group-based coordination and TTL-controlled multi-hop forwarding. Vehicles form groups led by a coordinator that manages resource allocation through control messages (Types A–D). A NACK from a group member triggers relaying by others, with linear combinations constructed using finite field arithmetic. TTL metadata determines how long a message should be forwarded, limiting redundancy.

However, setting TTLs in dense, 2D environments is challenging due to unpredictable vehicle movement. When transmission ranges (e.g., 200\,m) fall short of the simulation radius (e.g., 500\,m), coverage gaps lead to higher loss. Frequent group changes further introduce signaling delays. In contrast, our Adaptive MAC avoids grouping and uses local density estimation for lightweight, scalable relaying \cite{GNC-MAC}.

\subsection{Reliability-Bounded Network Coding}

Peng Wang et al. proposes a rate less coding-based broadcast scheme where a base station sends coded packets until a reliability threshold is met. The authors derive upper and lower bounds on the probability of successful decoding and estimate the number of transmissions needed to meet reliability constraints. This eliminates the need for ACKs and suits high-mobility scenarios.

We build upon this through adaptive, feedback-free broadcasting. Unlike the centralized BS model, we enable decentralized, bidirectional communication in urban VANETs, where each vehicle acts as both sender and receiver \cite{efficient_coding}.
\subsection{Game Theory in MEC-Assisted V2X Networks}
Haipeng Wang et al. propose an innovative computation offloading framework for Mobile Edge Computing (MEC) enhanced Vehicular-to-Everything (V2X) networks, mitigating vehicular computational limitations. The system allows vehicles to delegate intensive tasks to nearby peers (V2V) or network infrastructure (V2N). Crucially, a game-theoretic model underpins the offloading decision-making process, demonstrating convergence to a Nash equilibrium. This benefit ensures stable and efficient task distribution. The study introduces a computing offloading (CO) algorithm to minimize delay and energy, and an offloading-allocation (OA) algorithm to expedite convergence, both validated through simulations to optimize resource utilization and enhance V2X network performance \cite{Game-Theory}.
\subsection{RPCA: Distributed MAC with RSU Assistance}

 Zhang et al. apply a two-timescale MAC strategy: a large-scale phase for location-aware threshold updates and a small-scale phase for CSMA/CA-based contention. Upon winning access, a vehicle chooses between direct transmission, RSU-aided relaying, or re-contending, based on precomputed SNR thresholds derived from optimal stopping theory \cite{Coop_MAC_RSU_Distributed}.

While they achieve high throughput in unicast scenarios, their decision logic and RTS/CTS probing per vehicle pair are difficult to scale for broadcast-based VANETs. Our approach offers similar adaptivity with less complexity by focusing on many-to-many communication and statistical channel modeling.

\section{Proposed Approach}

Modern vehicular networks require reliable, low-latency communication for safety-critical applications, but they face significant challenges such as fluctuating network density, dynamic mobility, and wireless channel uncertainty. The number and positions of vehicles change rapidly, especially in urban environments, where signal reflections and irregular movement patterns are common due to buildings and other obstacles. Addressing these real-world dynamics requires adaptive mechanisms at both the MAC and PHY layers.

To address the above-mentioned gaps, we propose an Adaptive MAC strategy with density-aware enhancements and introduce RSUs to assist in message forwarding. We adopt a game theory-based approach at two levels to optimize the system. Game theory is used to have a varied beacon rate and also a varied transmission power at RSUs. The system is tested by simulating in radii of 250m and 500m, with a Manhattan-like mobility model to test it in urban regions. This ensures a more realistic scenario with varied vehicle density and street-like vehicle movement. This ensures the simulated data is based on realistic scenarios, unlike using idealistic assumptions mentioned in earlier papers. 
\\The core features of the proposed system include the following components:
\subsection{Density-Aware Contention Window Scaling}
Vehicles estimate the number of nearby nodes based on beacons received in the last second and adapt their minimum contention window (\(CW_{min}\)) accordingly. This allows high-density scenarios to reduce channel contention and avoid excessive collisions.

\subsection{Implicit Acknowledgment Mechanism}
Upon successful reception and subsequent forwarding of a vehicle's original beacon by an RSU, the originating vehicle may overhear this relay. If the relayed message content (specifically, the original sender ID and original message ID) matches a recently transmitted beacon by the overhearing vehicle, it infers successful delivery to the RSU. This positively acknowledges the sender's MAC layer, influencing retransmission decisions and contention window management without incurring the overhead of a dedicated ACK packet. Therefore, this mechanism relies on overhearing RSU-forwarded packets containing the vehicle's original message identifiers for confirmation.

\subsection{RSU-Assisted Relaying}
Four RSUs are placed at fixed positions within the circular area to opportunistically forward received vehicle messages. This aids in packet delivery when direct V2V links fail due to fading or congestion.
\subsection{Game Theory Based Transmission Energy}
Each RSU acts as a player in a non-cooperative game-theory model, and individually chooses the transmission energy based on the number of vehicles and optimizes the best energy to ensure that as many vehicles as possible receive the message, with minimal energy loss and interference. 
\subsection{Game Theory Based Adaptive Beacon Rate}
Dynamic beacon rate adaptation is achieved via a non-cooperative game-theoretic framework, integral to Decentralized Congestion Control (DCC). Individual vehicles autonomously optimize their transmission frequency, selecting from a discrete set to maximize a local utility function. This utility function critically balances the value of disseminating fresher situational data (favoring higher rates) against the cost of exacerbating channel congestion, a penalty modulated by beacon-derived local density estimates . Such adaptive control aims to enhance network scalability and ensure robust communication by curtailing packet collisions in dense vehicular environments while maintaining sufficient awareness.

These components collectively allow the proposed system to respond to fluctuating network conditions while maintaining low overhead and high scalability.

\section{Simulation Environment}
\label{sec:sim_env}
\begin{figure}[htbp]
    
    \centerline{\includegraphics[width=0.5\textwidth,keepaspectratio]{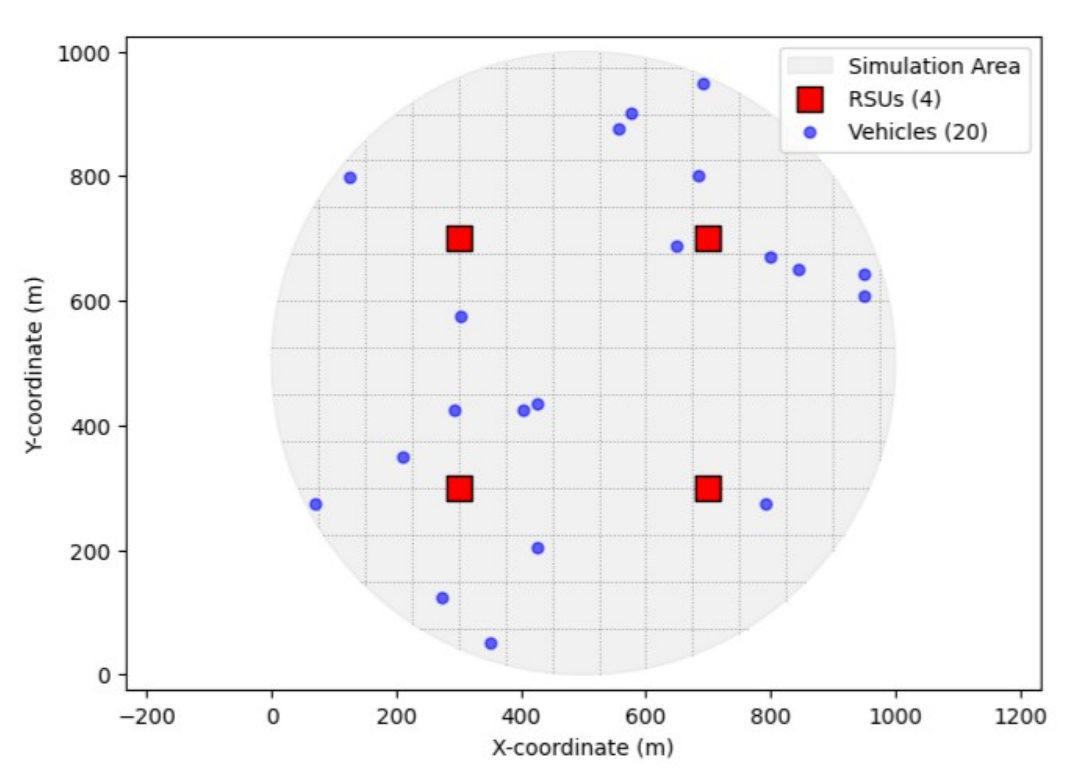}}
   \caption{Simulation layout displaying vehicles' initial position distributed in a manhattan-like grid circular area of 250\,m radius.}
    \label{fig:network_architecture}
\end{figure}

The simulation models a V2X network operating within a circular area of radius $R_{area} = \SI{250, 500}{\meter}$. Vehicles navigate this area using a Manhattan-like grid mobility model with a grid spacing of $G_{space} = \SI{60}{\meter}$ as shown in \ref{fig:network_architecture}. At intersections, vehicles probabilistically choose to turn or continue straight. A fixed number of Roadside Units (RSUs), $N_{RSU} = 4$, are strategically placed within the area to facilitate message relaying. The total simulation duration for each run is $T_{sim} = \SI{60}{\second}$.

\section{Communication Model}

\subsection{Physical Layer (PHY)}
Vehicles periodically broadcast beacon messages, with an initial/default interval of $T_{beacon\_default} = \SI{0.1}{\second}$ (10 Hz). Beacons have a fixed size $S_{beacon} = \SI{300}{bytes}$. The physical layer employs a data rate $R_{data} = \SI{6}{\mega bps}$. The channel bandwidth is $BW = \SI{1.08}{\mega\hertz}$. A PHY header overhead $T_{PHY\_ovh} = \SI{20}{\micro\second}$ is considered. The beacon transmission duration $T_{tx}$ is:
\begin{equation}
T_{tx} = \frac{S_{beacon} \cdot 8}{R_{data}} + T_{PHY\_ovh}
\label{eq:beacon_duration}
\end{equation}
The receiver noise floor, $\Nf$, in dBm is calculated as:
\begin{equation}
\Nf = \Nought + 10 \log_{10}(\BW) + \NF
\label{eq:noise_floor}
\end{equation}
where $\Nought = \SI{-174}{dBm/Hz}$ is the thermal noise power spectral density and $\NF = \SI{5}{dB}$ is the receiver noise figure.

\subsection{Channel Propagation Model}
A probabilistic Line-of-Sight (LOS)/Non-Line-of-Sight (NLOS) model is employed. Based on the distance $d$ between transmitter and receiver, the link is stochastically determined as LOS or NLOS, each associated with distinct path loss parameters. The path loss $\PL(d)$ in dB is modeled using a log-distance model:
\begin{equation}
\PL(d) = \PL(\dzero) + 10 \cdot \ple \cdot \log_{10}(d/\dzero) + X_{\sigma}
\label{eq:path_loss}
\end{equation}
where $\PL(\dzero) = \SI{47.85}{dB}$ is the free-space path loss at a reference distance $\dzero = \SI{1}{\meter}$ for a \SI{5.9}{\giga\hertz} carrier. The path loss exponent $\ple$ and shadowing standard deviation $X_{\sigma}$ are $\ple_{LOS} = 1.9$, $X_{\sigma,LOS} = \SI{2}{dB}$ for LOS, and $\ple_{NLOS} = 2.5$, $X_{\sigma,NLOS} = \SI{4}{dB}$ for NLOS.

The received power $\Prx$ in dBm from a transmitter with power $\Ptx$ in dBm is:
\begin{equation}
\Prx = \Ptx - \PL(d)
\label{eq:received_power}
\end{equation}

\subsection{Signal Reception Model}
A beacon is successfully decoded if its Signal-to-Noise Ratio (SNR) and Signal-to-Interference-plus-Noise Ratio (SIR) exceed predefined thresholds, $\SNRth = \SI{3.3}{dB}$ and $\SIRth = \SI{5.0}{dB}$, respectively.
\begin{align}
SNR &= \Prx - \Nf \ge \SNRth \label{eq:snr_check} \\
SIR &= 10 \log_{10}\left(\frac{P_{rx,mW}}{P_{noise,mW} + \sum P_{intf,mW}}\right) \ge \SIRth \label{eq:sir_check}
\end{align}
where $P_{rx,mW}$, $P_{noise,mW}$, and $P_{intf,mW}$ are powers in milliwatts.

\section{MAC Layer and Adaptive Mechanisms}
\label{sec:mac_adaptive}
The MAC layer is based on CSMA/CA principles. Vehicles employ density-aware contention window scaling.

\subsection{Density-Aware Contention Window Scaling (Vehicles)}
Vehicles estimate local density $\Nest$ by counting unique beacon senders successfully heard within a time window $T_{N_{est\_win}} = \SI{1.0}{\second}$. The minimum contention window $\CWmin'$ is adapted based on $\Nest$:
\begin{equation}
\CWmin' = \min\left(CW_{max}, CW_{min\_base} + \left\lfloor \frac{\Nest}{k_{N1}} \right\rfloor \cdot k_{N2}\right)
\label{eq:cw_adapt}
\end{equation}
where $CW_{min\_base}=15$, $CW_{max}=1023$, and $k_{N1}, k_{N2}$ are scaling factors ( $k_{N1} = 2.5, k_{N2} = 3$ ).

\subsection{Game-Theoretic RSU Transmit Power Adaptation}
RSUs adapt their transmit power $\Ptx^{RSU}$ using a non-cooperative game. Each RSU selects $\Ptx^{RSU}$ from a discrete set $\{P_1, ..., P_k\}$ to maximize its utility $U_{RSU}$, updated every $T_{RSU\_adapt} = \SI{0.5}{\second}$:
\begin{equation}
U_{RSU}(P_j) = w_{q} \cdot Q_{len} \cdot \frac{P_{j,mW}}{P_{def,mW}} - w_{p} \cdot \left(\frac{P_{j,mW}}{P_{def,mW}}\right)^2
\label{eq:rsu_utility}
\end{equation}
where $Q_{len}$ is the RSU's forwarding queue length, $P_{j,mW}$ is the candidate power in mW, $P_{def,mW}$ is a default RSU power in mW, and $w_q, w_p$ are weighting factors. Vehicle transmit power is constant at $\Ptx^{Veh} = \SI{20}{dBm}$.

\subsection{Game-Theoretic Vehicle Beacon Rate Adaptation}
Vehicles adapt their beacon interval $T_{beacon}$ (reciprocal of rate $R_{beacon}$) using a non-cooperative game. Each vehicle selects $R_{beacon}$ from a discrete set $\{R_1, ..., R_m\}$ to maximize $U_{Veh}$, updated every $T_{Veh\_adapt} = \SI{1.0}{\second}$:
\begin{equation}
U_{Veh}(R_j) = w_{f} \cdot R_j - w_{c} \cdot R_j \cdot \left(\frac{\Nest}{N_{scale}}\right)^{1.5}
\label{eq:vehicle_utility}
\end{equation}
where $N_{scale}$ is a reference density, and $w_f, w_c$ are weighting factors. The algorithm for this adaptation is shown in Algorithm~\ref{alg:beacon_rate_adapt}.
The key simulation parameters are summarized in Table~\ref{tab:sim_params}.

\begin{algorithm}
\caption{Vehicle Beacon Rate Adaptation}
\label{alg:beacon_rate_adapt}
\begin{algorithmic}[1]
\STATE \textbf{Input:} Vehicle $v$, Set of possible rates $R_{set}$, Current $_v$
\STATE \textbf{Output:} Updated $v.current\_beacon\_interval$
\STATE $max\_utility \gets -\infty$
\STATE $best\_interval \gets v.current\_beacon\_interval$
\FORALL{$rate_{option} \in R_{set}$}
\STATE $benefit \gets W_{freshness} \cdot rate_{option}$
\STATE $density\_factor \gets (N_{\mathrm{est},v} / N_{scale})$ \COMMENT{If $N_{scale}>0$}
\STATE $cost \gets W_{congestion} \cdot rate_{option} \cdot (density\_factor)^{1.5}$
\STATE $utility \gets benefit - cost$
\IF{$utility > max\_utility$}
\STATE $max\_utility \gets utility$
\STATE $best\_interval \gets 1.0 / rate_{option}$
    \ENDIF
\ENDFOR
\STATE $v.current\_beacon\_interval \gets best\_interval$
\end{algorithmic}
\end{algorithm}

\begin{table}[htbp]
\caption{Key Simulation Parameters}
\label{tab:sim_params}
\centering
\begin{tabular}{@{}lll@{}}
\toprule
Parameter                       & Symbol / Value                & Unit        \\ \midrule
\multicolumn{3}{@{}l}{\textbf{General Simulation Parameters}} \\
Simulation Time                 & $T_{sim}=60.0$                 & s           \\
Area Radius                     & $R_{area}=250,500$                & m           \\
RSU Count                       & $N_{RSU}=4$                   & -           \\
\multicolumn{3}{@{}l}{\textbf{PHY Layer Parameters}} \\
Vehicle Transmit Power          & $\Ptx^{Veh}=20$               & dBm         \\\\
RSU Init Transmit Power & $P_{\mathrm{tx,init}}^{\mathrm{RSU}} = 23$ & dBm          \\
RSU Power Levels                & $[18, 21, 23, 25, 27]$        & dBm         \\
Beacon Interval (Default)       & $T_{beacon\_default}=0.1$     & s           \\
Vehicle Beacon Rates            & $[2, 5, 10]$                  & Hz          \\
Bandwidth                       & $\BW=1.08$                    & MHz         \\
Data Rate                       & $R_{data}=6$                  & Mbps        \\
Beacon Size                     & $S_{beacon}=300$              & bytes       \\
PHY Overhead                    & $T_{PHY\_ovh}=20$             & µs          \\
Noise Figure                    & $\NF=5$                       & dB          \\
Thermal Noise Density           & $\Nought=-174$                & dBm/Hz      \\
\multicolumn{3}{@{}l}{\textbf{Channel Model Parameters}} \\
FSPL at \SI{1}{m} (\SI{5.9}{GHz}) & $\PL(\dzero)=47.85$         & dB          \\
Path Loss Exp. (NLOS)           & $\ple_{NLOS}=2.5$             & -           \\
Shadowing Std. Dev. (NLOS)      & $X_{\sigma,NLOS}=4$           & dB          \\
Path Loss Exp. (LOS)            & $\ple_{LOS}=1.9$              & -           \\
Shadowing Std. Dev. (LOS)       & $X_{\sigma,LOS}=2$            & dB          \\
SNR Threshold                   & $\SNRth=3.3$                  & dB          \\
SIR Threshold                   & $\SIRth=5.0$                  & dB          \\
\multicolumn{3}{@{}l}{\textbf{MAC Layer Parameters}} \\
Slot Time                       & $T_{slot}=10$                 & µs          \\
Min Contention Window (Base)    & $CW_{min\_base}=15$           & slots       \\
Max Contention Window           & $CW_{max}=1023$               & slots       \\
Max Retransmissions             & $N_{retx\_max}=3$             & -           \\
\multicolumn{3}{@{}l}{\textbf{Mobility and Grid Parameters}} \\
Grid Spacing                    & $G_{space}=60$                & m           \\
Min/Max Vehicle Speed           & $[5, 25]$                     & m/s         \\
Turn Probability (Straight)     & $P_{straight}=0.70$           & -           \\
\multicolumn{3}{@{}l}{\textbf{Game Theory Parameters (Veh. Rate Adapt.)}} \\
Adaptation Interval             & $T_{Veh\_adapt}=1.0$          & s           \\
Weight (Freshness)              & $w_f=1.0$                     & -           \\
Weight (Congestion)             & $w_c=0.2$                     & -           \\
Density Scaling Ref.            & $N_{scale}=15$                & vehicles    \\
\multicolumn{3}{@{}l}{\textbf{Game Theory Parameters (RSU Power Adapt.)}} \\
Adaptation Interval             & $T_{RSU\_adapt}=0.5$          & s           \\
Weight (Queue Benefit)          & $w_q=1.0$                     & -           \\
Weight (Power Cost)             & $w_p=0.5$                     & -           \\
\bottomrule
\end{tabular}
\end{table}

\begin{figure*}[t]
    
    \centering
    \begin{subfigure}[b]{0.35\textwidth}
        \includegraphics[width=\linewidth]{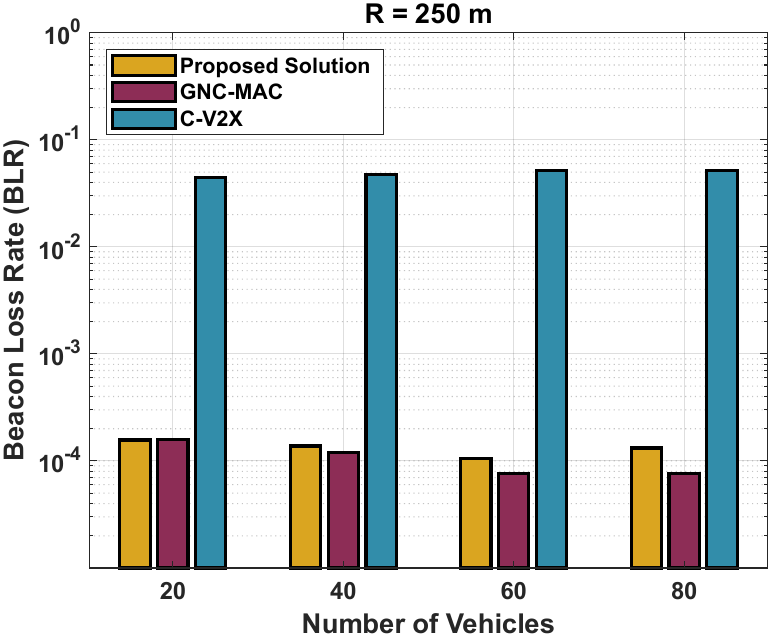}
        \caption{BLR at 250\,m radius.}
    \end{subfigure}
    \hspace{4em}
    \begin{subfigure}[b]{0.35\textwidth}
        \includegraphics[width=\linewidth]{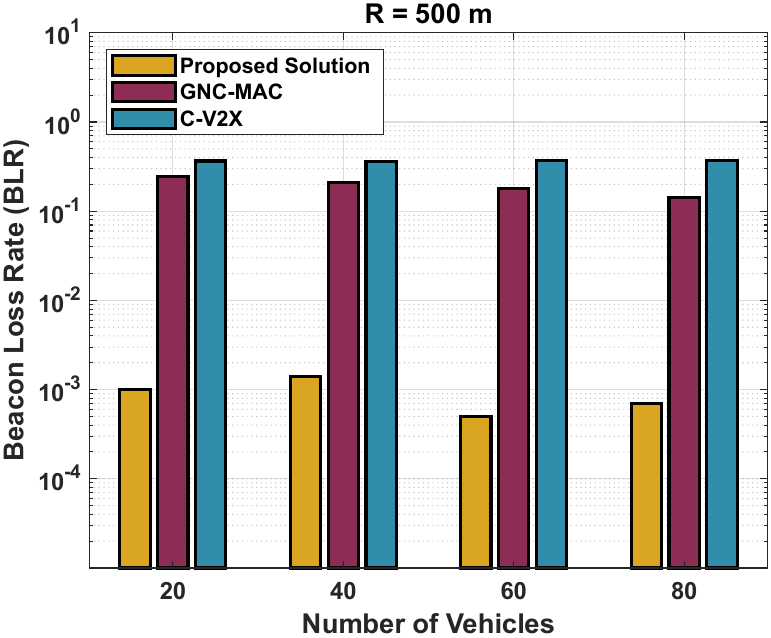}
        \caption{BLR at 500\,m radius.}
    \end{subfigure}

    \vspace{2em}

    \begin{subfigure}[b]{0.35\textwidth}
        \includegraphics[width=\linewidth]{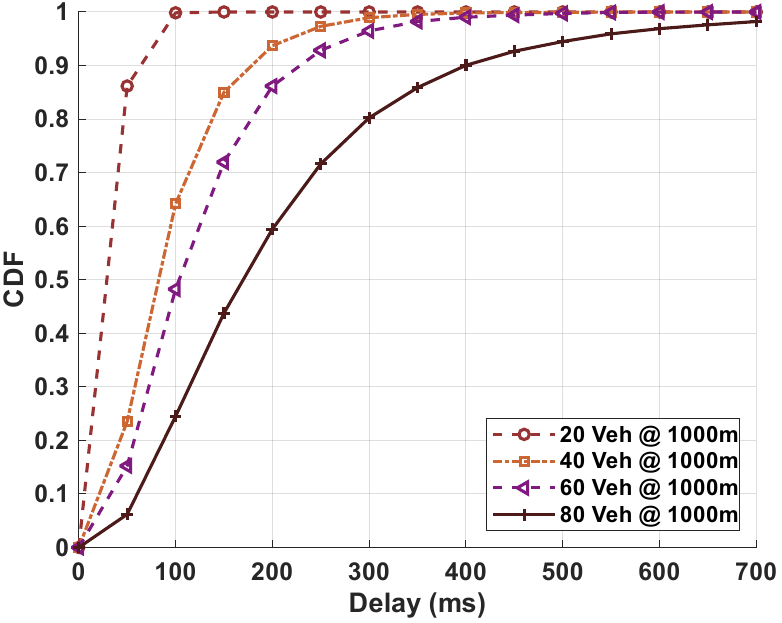}
        \caption{CDF vs Delay for 1000m radius}
    \end{subfigure}
    \hspace{4em}
    \begin{subfigure}[b]{0.35\textwidth}
        \includegraphics[width=\linewidth]{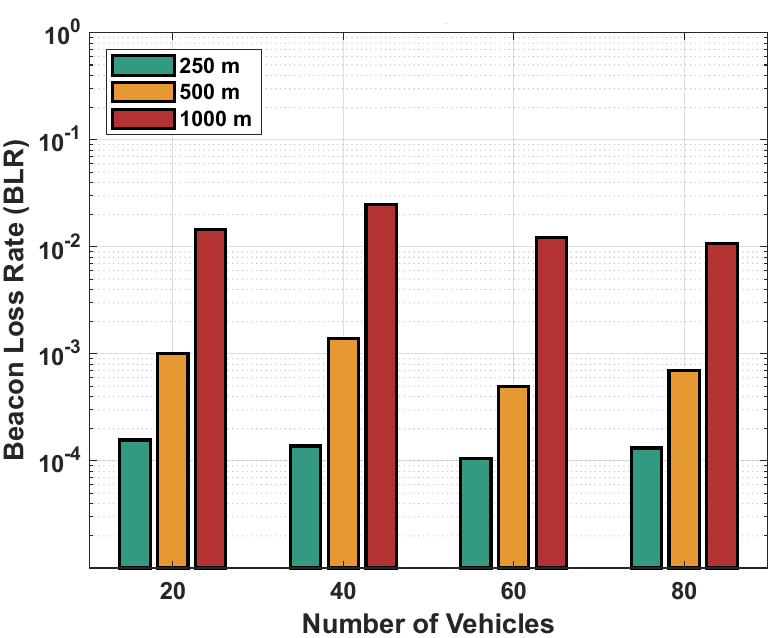}
        \caption{BLR at different radii}
    \end{subfigure}

    \caption{Performance comparison under varying vehicle densities and radii.}
    \label{fig:results}
\end{figure*}

\section{Results}

This section evaluates the proposed Adaptive MAC protocol in terms of Beacon Loss Rate (BLR) and transmission delay under varying vehicle densities. The performance is benchmarked against GNC-MAC and C-V2X. Simulations were conducted in circular areas with radii of 250 m, 500 m, and 1000 m, with vehicle counts ranging from 20 to 80 (Fig.~\ref{fig:results}). Four fixed RSUs were positioned as mentioned before to maximise the area covered, and node mobility followed grid-based movement. A beacon interval of 100 ms was chosen as the default value, and it was adapted based on \ref{alg:beacon_rate_adapt} throughout. Results for the proposed Adaptive MAC were obtained via simulation, while GNC-MAC and C-V2X data were extracted from \cite{GNC-MAC}.

The system was evaluated using the same SIR threshold (3.8 dB), as mentioned in GNC-MAC. However, the results obtained in GNC-MAC were based on an idealistic simulation, whereas the proposed system was tested with interference, realistic mobility to replicate high density and traffic conditions, and considerations for LOS and NLOS degradations. Hence, As shown in Fig.~\ref{fig:results}(a), (b), the proposed method performs comparably to GNC-MAC at 250 m and shows significant gains at larger radii. These gains stem from the use of implicit acknowledgments, RSU-based coordination, as well as game-theory-based adaptive beacon rates and transmission energy, in contrast to GNC-MAC’s multi-hop design. While GNC-MAC was tested in an interference-free setup with a 250 m radius and 200 m transmission range, the proposed Adaptive MAC method sustains robust performance despite realistic interference and losses, highlighting its practical scalability.

Although NC-MAC \cite{NC-MAC} is excluded from plots due to modeling differences (e.g., linear vehicle density), our approach achieved a lower BLR under comparable settings, reinforcing its reliability even without complex coding or centralized control.

Fig. \ref {fig:results}(c) demonstrates the delay performance of the proposed Adaptive MAC protocol under varying vehicle densities for 1000 m radius. This shows the reliability of our solution, with the CDF reaching 0.85 in less than 0.35 seconds. A similar pattern was seen for different radii.

As the communication radius increases to 1000 m, Fig.~\ref{fig:results}(d), the protocol continues to perform efficiently, even improving delay performance compared to smaller radii. Unlike traditional MAC schemes like NC-MAC, where delay increases due to denser contention and limited relaying, the proposed solution benefits from adaptive beacon and energy rates in larger areas, which reduces collisions and ensures messages can be delivered efficiently.

\section{Conclusion}
This paper presented a lightweight, scalable Adaptive MAC protocol for V2X systems. Simulations under realistic mobility, interference, and propagation conditions demonstrate consistently low BLR and low delay performance across various radii and vehicle densities. Unlike prior approaches that rely on centralized control or complex coding, our design leverages environment-aware adaptation and implicit acknowledgments to achieve high performance from 250 m to 1000 m, making it more feasible and realistic. The protocol shows strong potential for practical deployment in dense urban vehicular networks. The future scope would be hardware application and testing of the proposed approach. Furthermore, a game theory-based approach could be replaced by reinforced learning models to ensure longevity and better real-time tuning to ensure long-term stability of the system. Further improvements can be made to make the system energy aware by making the system multicasting-based and hence more robust in realistic traffic-based systems. Furthermore, lightweight encryption could be used to ensure RSUs are protected from overload and denial of service attacks.
\section{Acknowledgements}
We would like to thank Mars Rover Manipal, an interdisciplinary student team of MAHE, for providing the resources needed for this project.We also extend our gratitude to Dr. Ujjwal Verma for his guidance and support in our work.

\bibliographystyle{IEEEtran}
\bibliography{main} 
\end{document}